\documentclass[prb,twocolumn,showpacs,reprint,preprintnumbers,amsmath,amssymb]{revtex4}

\usepackage{color}
\usepackage{graphicx}
\usepackage{dcolumn}
\usepackage{bm}
\usepackage{graphicx}        
\usepackage{longtable}       
\usepackage{amsfonts}        
\usepackage{amsmath}         
\usepackage{amssymb}         
\usepackage{bm}              
\usepackage{color}           
\usepackage{dcolumn}         
\usepackage{epsfig}          
\usepackage{hyperref}        
\preprint{submitted to \PRB}

\newcommand{\bk}{\mathrm{\textbf{k}}}

\newcommand{\9}{\c{s}}

\newcommand{\ds}{\displaystyle}
\newcommand{\beq}{\begin{equation}}
\newcommand{\eeq}{\end{equation}}
\newcommand{\bea}{\begin{eqnarray}}
\newcommand{\eea}{\end{eqnarray}}

\newcommand{\AM}{Adv. Mater. }

\newcommand{\JAP}{J. Appl. Phys. }

\newcommand{\JPC}{J. Phys.: Condens. Matter }

\newcommand{\PR}{Phys. Rev. }

\newcommand{\PRB}{Phys. Rev. B }

\newcommand{\PRL}{Phys. Rev. Lett. }
\newcommand{\RMP}{Rev. Mod. Phys. }

\newcommand{\SSC}{Solid State Commun. }

\begin{document}

\title{Quasiparticle band structure of the almost-gapless transition-metal-based Heusler semiconductors}

\author{M. Tas$^1$}\email{murat.tas@kemerburgaz.edu.tr}
\author{E. \c{S}a\c{s}{\i}o\u{g}lu$^{2}$}\email{e.sasioglu@gmail.com}
\author{I. Galanakis$^3$}\email{galanakis@upatras.gr}
\author{C. Friedrich$^2$}
\author{S. Bl\"{u}gel$^2$}

\affiliation{$^1$Department of Basic Sciences,
\.{I}stanbul Kemerburgaz University, 34217 \.{I}stanbul, Turkey \\
$^2$Peter Gr\"{u}nberg Institut and Institute for Advanced
Simulation, Forschungszentrum J\"{u}lich and JARA, 52425
J\"{u}lich, Germany\\
$^3$Department of Materials Science, School of Natural Sciences,
University of Patras,  GR-26504 Patra, Greece}

\begin{abstract}

Transition-metal-based Heusler semiconductors are promising
materials for a variety of applications ranging from spintronics
to thermoelectricity. Employing the $GW$ approximation within the
framework of the FLAPW method, we study the quasi-particle band
structure of a number of such compounds being almost gapless
semiconductors. We find that in contrast to the
\textit{sp}-electron based semiconductors such as Si and GaAs, in
these systems the many-body corrections have a minimal effect on
the electronic band structure and the energy band gap increases by
less than 0.2~eV, which makes the starting point density
functional theory (DFT) a good approximation for the description
of electronic and optical properties of these materials.
Furthermore, the band gap can be tuned either by the variation of
the lattice parameter or by the substitution of the
\emph{sp}-chemical element.
\end{abstract}

\maketitle

\section{Introduction}

Gapless materials have been attracting remarkable attention after
the discovery of  graphene\cite{Novoselov} and topological
insulators\cite{Hasan} for applications in thermoelectricity,
nano-electronics, spintronics, and optics, due to their
high-mobility, robust transport and physical
properties.\cite{Tsidilkovski,Wang} They may have either a
quadratic energy dispersion like Hg$_{1-x}$Cd$_x$Te and
Hg$_{1-x}$Mn$_x$Te  or a linear energy dispersion like graphene
and organic salt $\alpha-$(BEDT-TTF)$_2$I$_3$. Materials with
linear dispersion obey the relativistic Dirac equation, and thus
have very high carrier mobilities. In gapless materials the charge
carriers can easily be excited from the valence band to the
conduction band without a cost of energy. Furthermore, one can
easily change their band structure by external perturbations like
pressure, temperature, chemical composition, electric and magnetic
fields. These innate properties make them desirable for
multifunctional tasks. For example, electrons and holes can be
fully spin-polarized simultaneously allowing tuneable spin
transport in spin-gapless semiconductors (SGSs),\cite{SGS} which
make them ideal spin injectors in spintronics devices.

Transition-metal (TM)-based gapless or almost-gapless
semiconductors crystallizing in half and full Heusler structure
display rich physics ranging from semimetallic and topological
insulator phases to heavy fermionic behavior.\cite{Graf} For
example, specific heat measurements of Heusler-type Fe$_2$VAl
showed that it is a candidate 3\textit{d} heavy-fermion
system.\cite{Nishino} Indeed, Guo \textit{et al.}\cite{Guo}
concluded via first-principles calculations that Fe$_2$VAl is a
nonmagnetic semimetal with a narrow pseudogap at the Fermi level.
On the other hand, narrow band-gap semiconductors in half Heusler
structure XNiSn (X = Ti, Zr, Hf) and CoTiSb are promising
materials for high-temperature thermoelectric applications with a
high figure of merit.\cite{Jaeger,Bos,Rausch}

Widely-spread standard first-principles calculations addressing
the electronic and optical properties of solid compounds including
also the TM-based gapless and almost-gapless  (also known as zero
or small band-gap, respectively) semiconductors (SCs) are based on
density-functional theory (DFT) within either the local density
approximation (LDA) or the generalized gradient approximation
(GGA) to the exchange correlation
functional.\cite{Guo,Weinert,Sato,Welle} As it is known, DFT is
restricted to the calculation of ground state properties, and
fails in describing the band gap and properties that rely on
excited states. The deficiency of DFT for materials from weak to
intermediate correlations can be healed by the $GW$ approximation,
which is shown to be quite successful in describing the excited
states, i.e., the band gaps and optical properties.\cite{Meinert}

The aim of the present work is a systematic study of the
electronic structure of the TM-based SCs [(XX$^{\prime}$)YZ, where
X, X$^\prime$, and Y are transition metal elements, and Z is an
\textit{sp} element] with almost zero band gaps. Due to the
presence of the TMs with narrow \textit{d}-bands the correlation
effects are expected to play a dominant role in electronic
structure of these materials. Using the $GW$ approximation within
the framework of the full potential linearized augmented planewave
(FLAPW) method, we show that in contrast to \textit{sp}-electron
based SCs such as Si and GaAs, in these systems the many-body
correlations have a minimal effect on the electronic band
structure. We find that for many compounds the change of the band
gap is less than 0.2~eV, which makes the starting point DFT-LDA
(DFT-GGA)) a good approximation for the description of electronic
and optical properties of these materials. Moreover, it is
demonstrated that the band gap can be tuned either by the
substitution of the Z element or by a variation of the lattice
parameter. The rest of the paper is organized as follows. In
Section\,\ref{section-2} we briefly describe the computational
scheme. Section\,\ref{section-3} presents the computational
results and discussion. In Section\,\ref{section-4} we give the
conclusions.

\section{Computational method}
\label{section-2}

The ground-state calculations are carried out using the FLAPW
method as implemented in the \texttt{FLEUR} code \cite{Fleur}
within the generalized gradient approximation (GGA) of the
exchange-correlation potential as parameterized by Perdew, Burke
and Ernzerhof (PBE).\cite{PBE} For all calculations we use angular
momentum and plane-wave cutoff parameters of $l_{\textrm{max}}=8$
inside the spheres and $k_{\textrm{max}}=4$~bohr$^{-1}$ for the
outside region. The DFT-PBE calculations are performed using a $16
\times 16 \times 16$  $\bk$-point grid. In order to accurately
describe the unoccupied states we use local orbitals. Note that in
the case of (CoCr)TiBi local orbitals are also used for semi-core
5$d$ states of bismuth.

We performed the one-shot $GW$ calculations using the Spex
code.\cite{SPEX} In the one-shot $GW$ approach off-diagonal
elements in the self-energy operator ${\ds
\Sigma_\sigma\left(E_{n\bk\sigma}\right)}$ are ignored and
corresponding expectation values of the local exchange-correlation
potential, ${\ds V_\sigma^{\textrm{XC}}}$, are subtracted in order
to prevent double counting. Within this framework, the Kohn-Sham
(KS) single-particle wavefunctions ${\ds
\varphi_{n\bk\sigma}^{\textrm{KS}}}$ are taken as approximations
to the quasiparticle (QP) wavefunctions. Hence, the QP energies
${\ds E_{n\bk\sigma}}$ are calculated as a first order
perturbation correction to the KS values ${\ds
E_{n\bk\sigma}^{\textrm{KS}}}$ as,\cite{Aguilera} \bea
E_{n\bk\sigma}=E_{n\bk\sigma}^{\textrm{KS}}+\left\langle
\varphi_{n\bk\sigma}^{\textrm{KS}}|
\Sigma_\sigma\left(E_{n\bk\sigma}\right)-V_\sigma^{\textrm{XC}}|\varphi_{n\bk\sigma}^{\textrm{KS}}\right\rangle,
\eea where $n$, $\bk$, and $\sigma$ are band index, Bloch vector,
and electron spin, respectively.

The dynamically screened Coulomb interaction $W$ is expanded in
the mixed product basis set  having contributions from the local
atom-centered muffin-tin spheres, and plane waves in the
interstitial region.\cite{Kotani} For mixed product basis set we
used the cutoff parameters ${\ds L_{\textrm{max}}=4}$ and ${\ds
G_{\textrm{max}}=3}$~Bohr$^{-1}$. For each compound three
different $\bk$-point grids are used to sample the full Brillouin
zone: $4 \times 4 \times 4$, $6 \times 6 \times 6$, and $8 \times
8 \times 8$. The relativistic corrections are treated at the
scalar-relativistic level (no spin-orbit coupling) for valence
states, while the full Dirac equation is employed for the core
states. We have converged the excited energies with the number of
unoccupied states and observed that with 300 bands the band gap is
converged to within less than 10~meV for all compounds under
study.

The considered  TM-based SCs  crystallize in the LiMgPdSn-type
crystal structure with the space group of F$\bar{4}$3m. They are
made up of four interpenetrating face-centered cubic (fcc)
lattices. The crystal structure is described by the chemical
formula (XX$^\prime$)YZ, where X, X$^\prime$, and Y are TMs, and Z
is an \textit{sp}-element. The atoms are placed at four
equidistant sites on the diagonal of the fcc lattice in sequence
of X-Y-X$^\prime$-Z. Valence of the TMs follows the same order as
in their chemical formula, i.e., X atoms have the highest valence,
while Y atoms have the lowest.

\begin{figure}
\begin{center}
\includegraphics[width=\columnwidth]{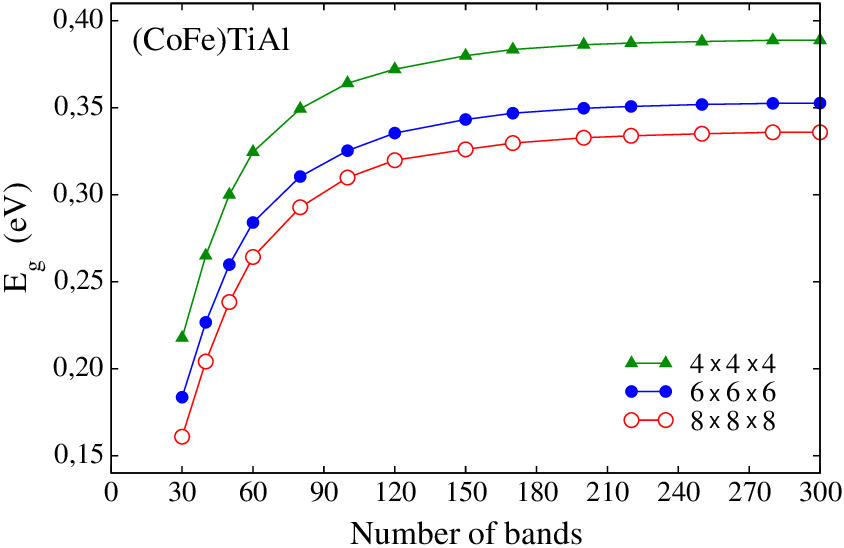}
\end{center}
\vspace*{-0.2cm} \caption{(Color online)  Convergence of the
energy band gap with the number of empty bands and the $\bk$-point
grid for (CoFe)TiAl.} \label{fig1}
\end{figure}

To identify TM-based gapless or almost-gapless SCs, gaps we use
the well-known Slater-Pauling (SP) rule\cite{SP,Galanakis} which
relates the total spin magnetic moment  per unit cell in
$\mu_\textrm{B}$ ($M_t$) of compounds to their total number of
valence electrons ($N_v)$. For the LiMnPdSn-type quaternary
compounds, the SP rule takes the form ${\ds
M_t=N_v-24}$,\cite{Ozdogan-JAP} and thus potential zero-gap SCs
should have 24 valence electrons per formula unit. We have chosen
as candidates the (FeMn)TiP, (CoMn)TiSi, (CoMn)VAl, (CoFe)TiAl,
(CoCr)TiP, (CoCr)TiAs, (CoCr)TiSb, and (CoCr)TiSb compounds. Since
$N_v=$24, the SP rule states that their total spin magnetic moment
should be zero.

\begin{table*}
\caption{Calculated equilibrium lattice constants $a$ (in {\AA}),
PBE and \textit{GW} band gaps, and transition energies (in eV)
between certain high-symmetry points for TM-based  SCs. \label{table}}
\begin{ruledtabular}
\begin{tabular}{llccrrrrrr}
& & & & \multicolumn{3}{c}{PBE} & \multicolumn{3}{c}{$GW$} \\
Compound & $a$(\AA) & E$_{\textrm{g}}^{\textrm{PBE}}$(eV) &
E$_{\textrm{g}}^{GW}$(eV) & $\Gamma \rightarrow \Gamma$ &
$\Gamma \rightarrow X$ & $X \rightarrow X$ & $\Gamma \rightarrow \Gamma$ & $\Gamma \rightarrow X$ & $X \rightarrow X$ \\
\hline
(FeMn)TiP   & 5.64 &  0.46 &  0.60 & 0.66 &  0.46 & 0.53 &  0.76 &  0.66 & 0.60  \\
(CoMn)TiSi  & 5.73 &  0.31 &  0.50 & 0.31 &  0.32 & 0.59 &  0.52 &  0.50 & 0.61  \\
(CoMn)VAl   & 5.74 & -0.07 & -0.15 & 0.18 & -0.07 & 0.25 &  0.26 & -0.15 & 0.02  \\
(CoFe)TiAl  & 5.81 &  0.08 &  0.30 & 0.08 &  0.12 & 0.45 &  0.30 &  0.33 & 1.08  \\
(CoCr)TiP   & 5.72 &  0.23 &  0.34 & 0.54 &  0.23 & 0.28 &  0.71 &  0.41 & 0.34  \\
(CoCr)TiAs  & 5.87 &  0.09 &  0.16 & 0.35 &  0.09 & 0.28 &  0.16 &  0.26 & 0.33  \\
(CoCr)TiSb  & 6.10 & -0.01 &  0.13 & 0.30 & -0.01 & 0.28 &  0.41 &  0.13 & 0.30  \\
(CoCr)TiBi  & 6.26 & -0.17 & -0.04 & 0.07 & -0.17 & 0.29 & -0.29 & -0.04 & 0.29  \\
\end{tabular}
\end{ruledtabular}
\end{table*}

\section{Results and discussion}
\label{section-3}

\subsection{DFT results}

We begin our discussion with the results obtained from the
standard GGA formalism of DFT. Since the compounds under study do
not exist experimentally, we have first determined the equilibrium
lattice constants using total energy calculations and fitting the
standard equation of state to the total energy curve. The obtained
results are presented in the first column of Table \ref{table}.
Lattice constants present large variations from 5.64 \AA\ for
(FeMn)TiP to 6.26 \AA\ for (CoCr)TiBi due to the large atomic
radius of the Bi atoms. Interestingly these values are similar to
the lattice constants of existing binary semiconductors.

In the case of Heusler compounds with 24 valence electrons, the
ground state can be either non-magnetic semiconducting or
fully-compensated ferrimagnetic semiconducting;\cite{Ozdogan-JAP}
in the latter case the atoms possess finite atomic spin magnetic
moments but the total magnetization is zero. Our first principles
calculations reveal that for the compounds under study the
non-magnetic semiconducting is the ground state since we were not
able to converge to a fully-compensated ferrimagnetic state
irrespectively of the initial configuration of the atomic spin
magnetic moments considered in our calculations.

\subsection{Band structure and energy gaps}

Using the DFT results as an input we performed one-shot \emph{GW}
calculations. Using (CoFe)TiAl as a representative example we will
discuss the changes in the band structure. The situation is
similar for the other compounds.  First we should establish the
convergence of the energy gap using \emph{GW} with respect to the
number of \textbf{k}-points and the number of energy bands taken
into account in the calculation. We present the results of our
convergence tests in Fig. \ref{fig1}. With a specific
\textbf{k}-point grid  we need around 300 empty bands to have
reliable values for the \emph{GW} energy gaps which corresponds to
about 75 empty bands per atom since there are 4 atoms per unit
cell, a value which is similar to the well-known semiconductors
like Si or Ge. As the grid becomes denser the energy gaps become
smaller but we can see that the difference between the
6$\times$6$\times$6 and 8$\times$8$\times$8 is of the order of
0.01-0.02 eV and thus the former can be safely used.

\begin{figure}
\begin{center}
\includegraphics[width=\columnwidth]{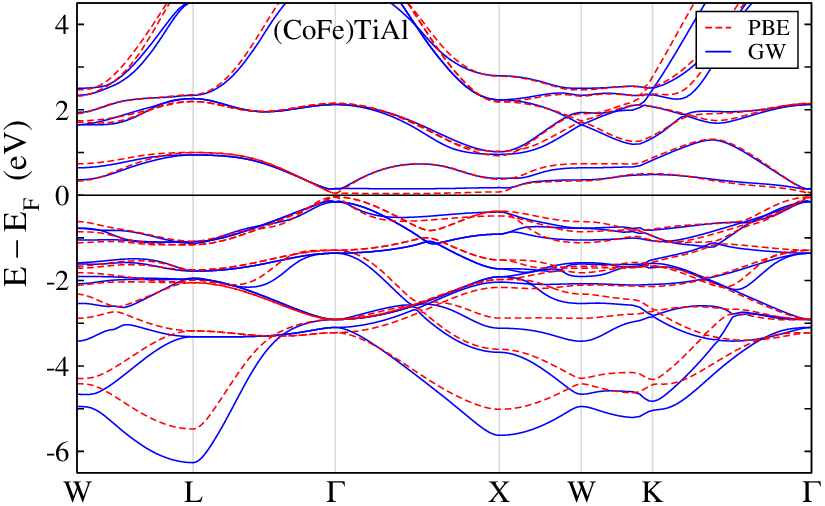}
\end{center}
\vspace*{-0.2cm} \caption{(Color online) Calculated electronic
band structure of (CoFe)TiAl along the high-symmetry directions in
the first Brillouin zone using either the PBE-GGA (red dashed
line) or the $GW$ (blue solid line) approximations.} \label{fig2}
\end{figure}

We present the calculated band structure in Fig. \ref{fig2} using
both the PBE parametrization of GGA and the \emph{GW} method.
First we shortly analyze the character of the bands at the
$\Gamma$ point, giving the character of the orbitals in real
space. The arguments are similar to the case of usual
half-metallic full and ordered quaternary Heusler
compounds.\cite{Galanakis,Ozdogan-JAP} There is an \emph{s}-band
low in energy not shown in Fig. \ref{fig1}. The lowest shown bands
are the Al \emph{p}-bands, which are triple-degenerate at the
$\Gamma$-point and which also accommodate the \emph{d}-charge of
the transition metal atoms. Then there are the double-degenerate
$e_g$ and triple-degenerate (at the $\Gamma$-point) $t_{2g}$
\emph{d}-hybrids of the transition metal atoms which are in
tetrahedral symmetry. The states just below and just above the
Fermi level are the triple-degenerate $t_{1u}$ and the
double-degenerate $e_u$ states which are located exclusively at
the Co and Fe sites in octahedral symmetry; for an extensive
discussion see Refs. \onlinecite{Galanakis} and
\onlinecite{Ozdogan-JAP}.

The band structure using \textit{GW} presented in Fig \ref{fig2}
looks qualitatively similar to the PBE-GGA results. The bands,
however, differ quantitatively especially at energies away from
the Fermi energy where the $p$ and the bonding $e_g$ and $t_{2g}$
hybrids reside. The change in the bands originating from the $e_u$
and $t_{1u}$ hybrids is much smaller and less visible. Mainly
\textit{GW} pushes the conduction ($e_u$ hybrids) and valence
($t_{1u}$ hybrids) bands, respectively up and down, hence opening
the band gap, but the broadening does not exceed 0.2 eV. As a
result within PBE-GGA, (CoFe)TiAl is an almost-gapless
semiconductor with a direct gap at the $\Gamma$-point of 0.08 eV
while within \textit{GW} one could classify it as a narrow-band
semiconductor with a direct energy gap of 0.30 eV.

Finally in Table\,\ref{table} we present both the PBE-GGA and $GW$
energy gaps for all compounds under study. First, we should note
for any chosen \textbf{k}-point the bands are well-separated by a
finite energy gap. This is reflected in the direct $\Gamma
\rightarrow \Gamma$ and $X \rightarrow X$ energy band gaps
presented in the table. For many of the compounds under study; the
fundamental energy gaps (E$_{\textrm{g}}^{\textrm{PBE}}$ and
E$_{\textrm{g}}^{GW}$) correspond to the indirect $\Gamma
\rightarrow X$ gaps as  seen in Table\,\ref{table}. Moreover in
the case of (CoMn)VAl, (CoCr)TiSb and (CoCr)TiBi, the overlap of
the valence and conduction bands is such that instead of getting a
semiconductor we get a semimetallic-like behavior of the density
of states, \textit{i.e.} in this case the bottom of the conduction
band is lower in energy than the top of the valence band. This
behavior is reflected in the negative values given in the table.

The size of the gaps increases with the valence of the $sp$ atom
as we move from Al to Si and then to P where the valence electrons
increase by one. A similar remark can be made when we move along a
column of isovalent chemical elements from the lighter P to As, Sb
and finally Bi. To understand this trend we have to remark that
the $d$-hybrids below the Fermi level have also a small portion of
$p$ character. For the lighter element like Al this admixture is
sizeable and contributes to the opening of the gap similarly to
the usual $p-d$ hybridization scheme in
semiconductors.\cite{Weinert} As Al is substituted by heavier
atoms, their $p$ states are located deeper in energy, the
admixture of the p character in the $d$ bands decreases and the
energy gap shrinks. The use of \textit{GW} increases
systematically the values of both direct and indirect bands gaps
but this increase does not exceed 0.2 eV in all cases. Even this
moderate increase can transform the character of (CoCr)TiSb from
semimetallic-like to semiconducting.

\begin{figure}[!t]
\begin{center}
\includegraphics[width=\columnwidth]{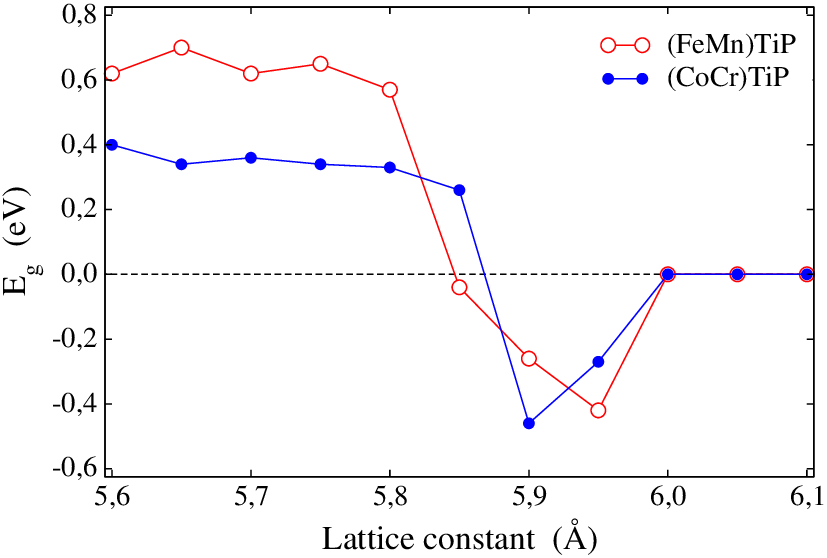}
\end{center}
\vspace*{-0.2cm} \caption{(Color online) Energy band-gap tuning
with the variation of the lattice parameter for the compounds
(FeMn)TiP and (CoCr)TiP using the $GW$ approximation.}
\label{fig3}
\end{figure}

Finally, we should also discuss the applicability of \emph{GW} to
the study of the electronic band structure of materials. Although
in principle \emph{GW} is a more elaborate method to study the
band structure of solids since it includes the quasiparticle
corrections, its success with respect to standard density
functional calculations is not guaranteed and results should be
confirmed by experiments. Recently, Meinert \textit{et
al}.\cite{Meinert} studied the band structure of the half-metallic
ferromagnetic full Heusler compounds, Co$_2$MnSi and Co$_2$FeSi,
via one-shot $GW$ calculations and found that the many-body
corrections are crucial especially for the latter system, and
experimentally studied spectra of both materials are well
described by the $GW$ calculations. In contrast, the band gap of
wurtzite ZnO is repeatedly underestimated by more than 1~eV within
the $GW$ calculations with respect to its experimental value. A
recent $GW$ calculation with the FLAPW method has found its band
gap higher than previous studies, but still lower than its
measured value.\cite{Friedrich-ZnO} Better agreement with
experiment can be achieved with a self-consistent
procedure.\cite{Klimes} Similarly, in the case of FeS$_2$ pyrite,
on the other hand, the one-shot $GW$ calculations find a lower
value for the fundamental band gap.\cite{Schena}

\subsection{Effect of the lattice parameter}

Expanding or compressing lattices, and even slight changes in the
chemical composition alter dramatically the lattice constants of
compounds and their band-gap energies. This so-called band-gap
tuning, or band-gap engineering, by any means  has become quite
important in designing new multifunctional devices. The $GW$
band-gap energies of compounds (FeMn)TiP and (CoCr)TiP are plotted
as a function of their lattice constant in Fig.\,\ref{fig3}.
(FeMn)TiP is found to be a SC in both PBE and $GW$ calculations at
its equilibrium lattice constant of 5.64~\AA ~(see
Table\,\ref{table}), but it is predicted to be a semimetal when
its lattice constant is increased just about 3.6\%. Similarly,
(CoCr)TiP seems to make a sharp transition from a semiconducting
to a semimetallic phase when its lattice constant is extended
approximately 2.3\% from its equilibrium value. Both compounds are
found to be gapless when their lattice constants are assumed to be
6~\AA. Tailoring of the band-gap energy by substitution of
\textit{sp}-elements in compounds (CoCr)TiZ is already discussed
above.

\section{Conclusions}
\label{section-4}

We have studied the quasi-particle band structure of several
transition-metal-based almost-gapless Heusler semiconductors
(XX$^\prime$)YZ by using the one-shot $GW$ approximation within
the framework of the FLAPW method. We find that in contrast to
\textit{sp}-electron based semiconductors such as Si and GaAs, in
these systems the many-body correlations have a minimal effect on
the electronic band structure. For these compounds the change of
the energy band gap is less than 0.2~eV. Thus, the standard
density-functional-theory based first-principles calculations are
a good approximation for describing the electronic properties of
these materials. Furthermore, the band gap of these compounds can
be tuned by varying the lattice parameter or by substituting the
main group \emph{sp}-element.

\begin{acknowledgments}
M. Tas acknowledges kind hospitality of the Peter Gr\2nberg Institut and Institute for Advanced Simulation,
Forschungszentrum J\2lich, Germany.
\end{acknowledgments}

\end{document}